\documentclass[twocolumn,           
               showpacs,            
               nopreprintnumbers,     
               aps,                 
               prd,          	    
               letterpaper,             
               nofootinbib,         
               tightenlines,        
               floats,floatfix,      
               showkeys
               ]{revtex4-1}

\usepackage{graphicx}
\usepackage{dcolumn}
\usepackage{bm}
\usepackage{amsmath}
\usepackage{amsfonts,amssymb}
\usepackage{color}
\usepackage{enumerate}
\usepackage{float}
\usepackage{subfig}

\begin{document}

\title{Testing noncommutativity-like model as a galactic density profile}

\author{Juan J. Ancona-Flores$^{1}$}
\email{jancona16@alumnos.uaq.mx}

\author{A. Hern\'andez-Almada$^{1}$}
\email{ahalmada@uaq.mx}

\author{Miguel A. Garc\'{\i}a-Aspeitia$^{2,3}$} 
\email{aspeitia@fisica.uaz.edu.mx}

\affiliation{$^{1}$Facultad de Ingenier\'ia, Universidad Aut\'onoma de Quer\'etaro, Centro Universitario Cerro de las Campanas, 76010, Santiago de Quer\'etaro, M\'exico.}

\affiliation{$^2$Unidad Acad\'emica de F\'isica, Universidad Aut\'onoma de Zacatecas, Calzada Solidaridad esquina con Paseo a la Bufa S/N C.P. 98060, Zacatecas, M\'exico.}

\affiliation{$^3$Consejo Nacional de Ciencia y Tecnolog\'ia, Av. Insurgentes Sur 1582. Colonia Cr\'edito Constructor, Del. Benito Ju\'arez C.P. 03940, Ciudad de M\'exico. M\'exico.}

\begin{abstract}
Noncommutative-like model (NC-like) is an interesting alternative inspired by string theory to understand and describe the velocity rotation curves of galaxies without the inclusion of dark matter particles. In a natural way, a Gaussian density profile emerges and is characterized by a parameter $\theta$, called the NC-like parameter. Hence we aim to confront the NC-like model with a galaxy sample of the SPARC catalogue to constrain the model parameters and compare statistically with the Einasto density profile using the Akaike and Bayesian information criteria. According to our results, some galaxies prefer the NC-like over the Einasto model while others do not support NC-like.
\end{abstract}

\date{\today}
\maketitle

\section{Introduction}	

Nowadays, dark matter (DM) is one of the most elusive components in modern astrophysics and cosmology, observed in several phenomena, from rotation curves in galaxy dynamics \cite{Rubin2001} to large scale structure in the Universe \cite{,FW,Diaferio:2008jy,Aghanim:2018}. Additionally, DM is confirmed in the Cosmic Microwave Background Radiation (CMB) \cite{Aghanim:2018}, having percentages around $\sim29\%$ of the total components. Moreover, the addition of a cold dark matter component and its analysis with computational simulations is in agreement with the knowledge of structure formation at large scales and the observed distribution of structures \cite{Springel:2005nw,Sawala:2015cdf}.

The assumption of a DM halo generates the stability of the galactic structure and a compatibility with the observed velocity rotation at large radius. Thus, the corresponding rotation velocities of the galaxies can be described by several empirical density profiles, based for example, on N-body simulations like the Navarro-Frenk-White (NFW) \cite{NFW} profile or by  phenomenological models such as pseudo isothermal (PISO) \cite{piso}, Burkert \cite{Burkert}, Einasto \cite{Einasto}, scalar field dark matter (SFDM) \cite{Hernandez-Almada:2017mtm}, among other \cite{Dehnen}. However, each of these models has its advantages and disadvantages\footnote{See for example in Ref.  \cite{Navarro1997,*Subramanian2000,*Salucci}, where the halo profiles are more dense and more pronounced than those inferred observationally.}, proving that there is not yet a definitive model.

Despite the amount of empirical models proposed during the last years, the microscopic nature of DM is still a mystery. In literature, many mechanisms have been proposed in order to explain DM and its possible relation with the density profiles in galaxies and also with the large scale structure. For instance, supersymmetric models related to weak interacting massive particles (WIMPS) are the most accepted candidates by the scientific community to explain DM, due to their advantages in the standard model of particles (SM) or in quantum gravity like string theory \cite{Horava:1996ma,*Horava:1995qa}. However, several interesting alternatives have emerged, for example: SFDM as in the case of axions \cite{Lee:1995af,Barranco:2010ib} or ultralight scalar fields \cite{UrenaLopez:2000aj,*doi:10.1063/1.4748537,*Robles:2014ysa}; or even extensions to General Relativity (GR) like $f(R)$ theories \cite{Martins/Salucci:2007}, brane-world models \cite{Randall-I,Randall-II,mk,Garcia-Aspeitia:2015eja,Garcia-Aspeitia:2018fvw,Garcia-Aspeitia:2016kak}, etc.

Another possibility to explain the rotation curves of galaxies comes from noncommutativity (NC) models \cite{Rahaman:2010vv}. The idea emerge from string theory, based on the assumption of NC space-time coordinates, obtaining a new type of gauge theory via the Seiberg-Witten map \cite{Seiberg}. In this sense, one of the most common examples of NC comes from quantum mechanics in two dimensions in where are encoded the new commutation relations via the coordinate operators. Therefore, many studies have been done in order to constrain the NC parameter, obtaining values approximately equal to the Planck length \cite{Romero} or even Trans-Planckian. Hence, based on these hypotheses, a Gaussian distribution of minimal width may be used instead of a Dirac-delta function\footnote{The Dirac-delta function help to describe a point-like structure in the standard case of quantum field theory.}. In fact, this change is also motivated when the amplitude between two states with different mean position is estimated using the Feynman path integral \cite{Smailagic:2003yb,*0305-44}.

On the other hand, as the NC modifies the energy-momentum tensor presented in GR, the smeared objects described by a Gaussian distribution may be used to study macroscopic systems such as black holes \cite{Nicolini:2008aj,*Spallucci:2009zz,*Ansoldi:2006vg}, in which is presented a form of NC model to alleviate singularities\footnote{In this context it is shown how the Ricci scalar at zero radius, is a function of the NC-like parameter in the form $R(0)\propto\theta^{-3/2}$ \cite{Nicolini:2008aj}, presenting no divergences.} as well as issues of galactic dynamics (rotation curves) \cite{Rahaman:2010vv}, the latter of which will be the focus of our study.

In this sense, inspired by the study of Rahaman et al. \cite{Rahaman:2010vv}, where they suggest that a density profile inspired by NC could produce the same intra-galactic dynamics as a DM profile (called hereafter NC-like). The idea behind this is as follows: assume the DM halo is replaced by a Gaussian density profile, arising from NC effects smearing out the density of a central compact body. This new profile is then parameterized by a central density as well as a length-scale that emerges from NC effects at the quantum level. In fact, this model could be expected to provide a good fit, as its functional form could be reproduced by an Einasto density profile in the appropriate limit. Moreover, because the smoothness shape of a Gaussian around the central region, the study of the NC-like model is also motivated by the results found in \cite{Li_2020} which suggest that the central density of the galaxies flattens out, forming a core. In this vein, we propose a robust statistical study through the current Spitzer Photometry and Accurate Rotation Curves (SPARC) sample \cite{SPARC:2016} by performing a Markov Chain Monte Carlo (MCMC) analysis. In particular, we constrain the associated free parameters $\theta$ and improve the statistical test using Akaike and Bayesian information criteria\footnote{There are other alternatives to compare models statistically such as Bridge Criterion \cite{Ding:2018}.} to compare the NC-like model with the Einasto density profile due to its similarity with the NC-like density. 

We organize this paper as follows: In Sec. \ref{NC} we present the rotation velocity of NC-like and Einasto, through the density profile associated for each model. In Sec. \ref{Res} we present the Bayesian MCMC analysis of NC-like and Einasto performed through a galaxy sample provided by SPARC. Finally, in Sec. \ref{Con} we discuss the results obtained and the viability of using NC in other astrophysical or cosmological studies.

In what follows, we work in units in which $c=\hbar=1$, unless explicitly written.

\section{Noncommutative-like  and Einasto rotation velocity} \label{NC}

This section presents the DM profiles associated to NC theory and the phenomenological Einasto density profile with the aim to confront them with observational rotation curves of galaxies.

We start considering $V$ as the sum of the velocities of the disk ($V_{\rm disk}$), gas ($V_{gas}$), and the halo ($V_{\rm halo}$). In other words
\begin{equation}\label{eq:Vtotal}
V^2(r)=\Upsilon_{\rm disk}V_{\mathrm{disk}}^2 + V_{\mathrm{gas}}^2 + V_{\mathrm{halo}}^2\,,
\end{equation}
where $\Upsilon_{\rm disk}$ is the stellar mass-to-light ratio, which is in general a function of $r$ and in this work it is considered as a free parameter.  While the first two terms are given by observations, the latter will be described by the two density profiles studied below.

Hence, the rotation velocity at Newtonian level is related to the effective 
potential and is given for the halo as
\begin{equation} \label{rotvel}
V^2(r)_{halo}=r\left\vert \frac{d\Phi(r)}{dr}\right\vert=\frac{G\mathcal{M}(r)_{halo}}{r}\,, 
\end{equation}
where $\mathcal{M}(r)_{halo}$ is its mass within a radius $r$, obtained through the integral
\begin{equation}
    \mathcal{M}(r)_{halo}=4\pi\int_0^r\rho_i(r')r'^{2}dr', \label{Mass}
\end{equation}
where $\rho_i(r')$ is associated to the halo density profile, where in our case will be for NC-like and Einasto profiles.

\subsection{Noncommutativity-like density profile} \label{NoncomT}

For the NC-like case, the density profile is given by \cite{Smailagic,Nicolini:2005vd}
\begin{equation}\label{Gauss}
\rho(r)_{\rm NC}=\rho_0\exp\left(-\frac{r^{2}}{4\theta}\right)\,, 
\end{equation}
where $\rho_0$ is the central density at $r=0$, not presenting divergences\footnote{Is in this vein, that it is not necessary, extend the region in the form $r-R_0$ as Ref. \cite{Rahaman:2010vv} suggest.} and it is defined as
$\rho_0=M/(4\pi\theta)^{3/2}$,  being $M$ the halo mass and $\sqrt{\theta}$ is a characteristic length of the model, being both $\rho_0$ and $\sqrt{\theta}$ the free parameters which will be constrained by observations and it is assumed that $\sqrt{\theta}$ is a constant term\footnote{A variable $\sqrt{\theta}$ function, implies additional hypotheses that generates unnecessary complications and interpretations.}.
We remark that the density profile presented in Eq. \eqref{Gauss} is inspired by NC, expecting that $\theta$ is constrained to be of the order of kiloparsec, being an emergent variable of noncommutative microscopic quantities.

Hence, the NC-like velocity rotation is expressed in the following way through equations \eqref{rotvel} and \eqref{Mass} as
 
\begin{equation}\label{velrotnon}
V^{2}_{\rm NC}(r)=4\pi G\rho_0\frac{\theta^{3/2}}{r}\Big\vert \sqrt{\pi}{\rm Erf}\left( \frac{r}{2\sqrt{\theta}}\right)- \frac{r}{\sqrt{\theta}}\exp\left(- \frac{r^2}{4\theta}\right)\Big\vert, 
\end{equation}
where ${\rm Erf}(x)$ is the 
error function.

\subsection{Einasto density profile} \label{Ein}

As it was mentioned before, the NC-like model is a particular case of Einasto's model given by \cite{Einasto1}
\begin{equation} 
\rho_{\rm E}(r)=\rho_{-2}\exp\left\{-2n\left[\left(\frac{r}{r_{-2}}\right)^{1/
n}-1\right]\right\}.
\end{equation} 
The $r_{-2}$ is 
the radius where the density profile has a slope $-2$ and $\rho_{-2}$ is the 
local density at that radius; the parameter $n$ is known as Einasto index which 
describes the shape of the density profile.

From Eqs. \eqref{rotvel} and \eqref{Mass}, the following form of the rotation velocity is
\begin{eqnarray}
V_{{\rm E}}^2(r)&=&4\pi Gnr^2_s\rho_{-2}\exp(2n)(2n)^{-3n}\left(\frac{r_s}{r}\right)\nonumber\\&&\times\gamma\left(3n,2n\left(\frac{r}{r_{-2}}\right)^{1/n}\right), \label{Evelrot}
\end{eqnarray}
where
\begin{equation}
 \gamma(a,x)=\int_{0}^{x}t^{a-1}e^{-t}dt, 
\end{equation} 
is the incomplete gamma function. Notice that when $n=0.5$ we recover the functional form of Eq. \eqref{velrotnon}. 

In order to compare both profiles, it is convenient to compare the densities at a same point, hence the NC density in the core $\rho_0$ is related to $\rho_{-2}e^{2n}$ in Einasto case, and the smear parameter $\sqrt{\theta}$ corresponds to $r_{-2}/2$. One disadvantage of Einasto profile is that have more free parameters than NC-like model which implies a better fit with galaxies rotation curves.

\section{Data samples and fits} \label{Res}

In this section we describe the procedure to constrain phase-space of the model parameter 
using a rotation curve sample collected in the SPARC catalogue \cite{SPARC:2016}.
We model the RC distribution of the galaxies
as the sum of the stellar disk, gas   
and a spherical dark component. Both disk and gas component
are provided by this catalogue and the latter are
the NC-like or Einasto's models presented in Eqs. \eqref{velrotnon} and \eqref{Evelrot} respectively.

To test both DM models NC-like and Einasto, we select a subset of nine of the new general catalogue (NGC) galaxies of low surface brightness (LSB) listed in Table \ref{tab:ResultsNC} which satisfy the following conditions: the galaxy contains at least 10 data points to avoid an overfitting, the last point must be measured at $r>5$kpc as a measurement of the galaxy size and the galaxies do not contain a bulk component as this component affects mainly the central region\footnote{The elected subsample of galaxies does not contains any special relation that could benefit one or another theory.}. 
Based on MCMC method implemented in {\it lmfit} package \cite{lmfit:2016}, after initially discarding 400 chains (burn-in) to stabilize the steps, a total of $10,000$ chains is generated to explore the confidence region of the parameter space, ${\bf\Theta}=(\rho_0, \sqrt{\theta}, \Upsilon_d)$ and $(\rho_{-2}, r_{-2}, n, \Upsilon_d)$ for NC-like and Einasto respectively. Additionally, we consider flat priors on the region: densities ($\rho_0$ or $\rho_{-2}$): $[10^{5},10^{10}] \, {\rm M}_\odot/{\rm kpc}^3$, radius ($\sqrt{\theta}$ or $r_{-2}$):$[0.5, 30]$kpc, and $\Upsilon_{\rm disk}:[0,1]$.
The best fit values of the parameters are obtained by maximizing the likelihood function 
$\mathcal{L}({\bf\Theta})\, \propto \, \mathrm{exp}[-\chi^2({\bf\Theta})/2]$, where
\begin{equation}\label{eq:chi2}
\chi^2({\bf\Theta})=\sum_i^N \left( \frac{V_{\mathrm{obs}}^i-V({\bf\Theta})}{dV_{\mathrm{obs}}
^i} \right)^2\,.
\end{equation}
In the above expression, $V_{\mathrm{obs}}^i\pm dV_{\mathrm{obs}}^i$ is the observed velocity and its corresponding 
uncertainty at the radial distance $r_i$ and $V({\bf\Theta})$ is the theoretical velocity.

\begin{figure*}[h]
   \centering
   \subfloat[]{
        \label{fig:fit1}         
        \includegraphics[width=0.3\textwidth]{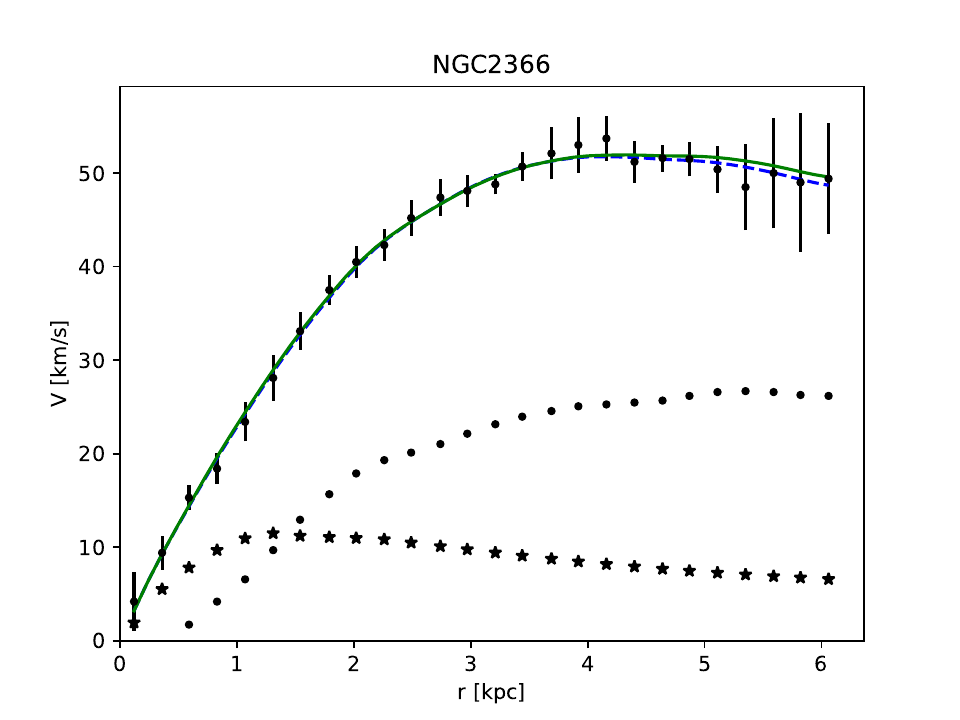} }           
   \subfloat[]{
       \label{fig:fit2}         
        \includegraphics[width=0.3\textwidth]{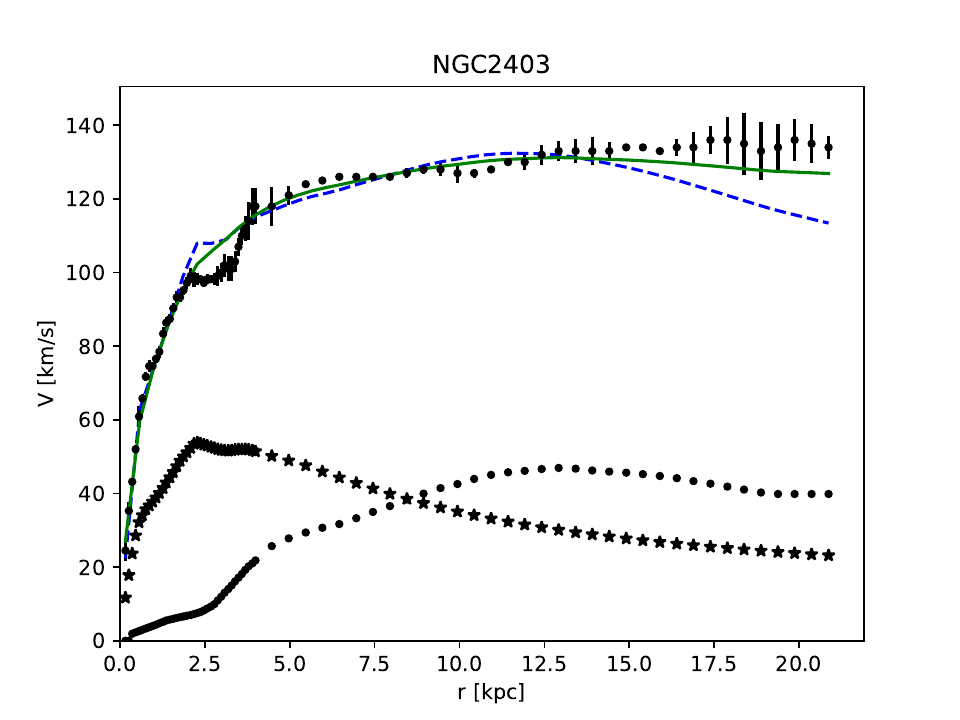} }
   \subfloat[]{
        \label{fig:fit3}         
        \includegraphics[width=0.3\textwidth]{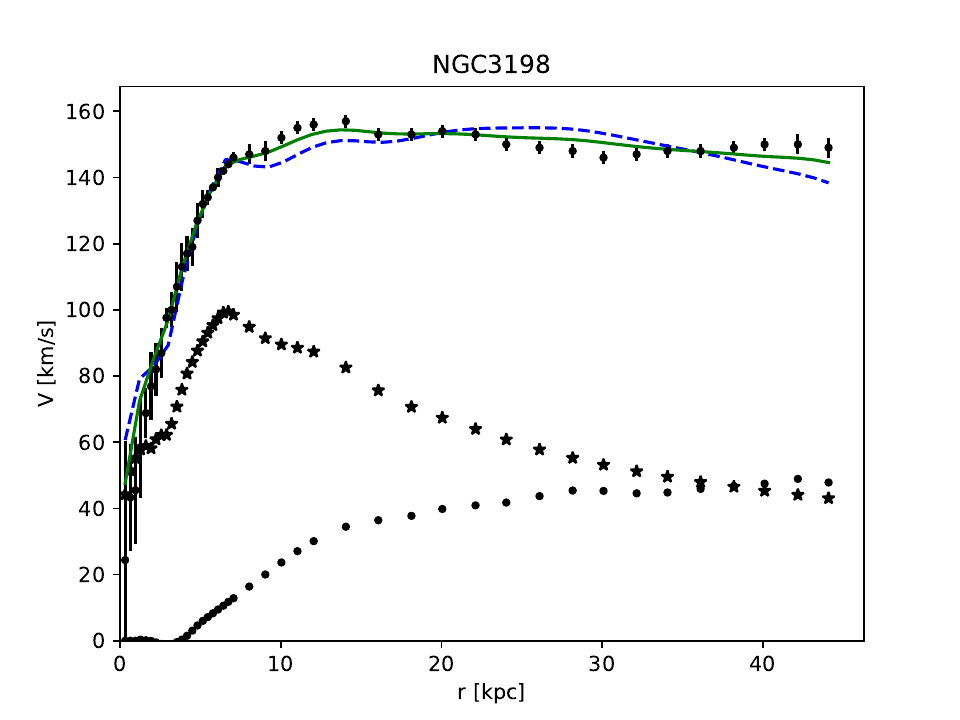} } \\
   \subfloat[]{
        \label{fig:fit4}         
        \includegraphics[width=0.3\textwidth]{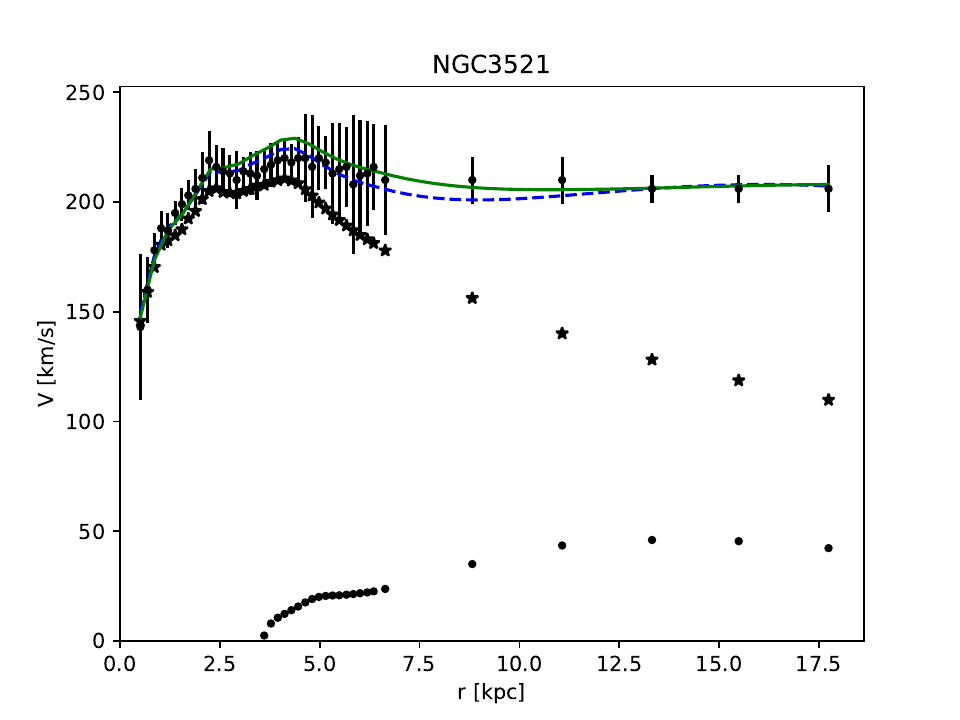} } 
   \subfloat[]{
        \label{fig:fit5}         
        \includegraphics[width=0.3\textwidth]{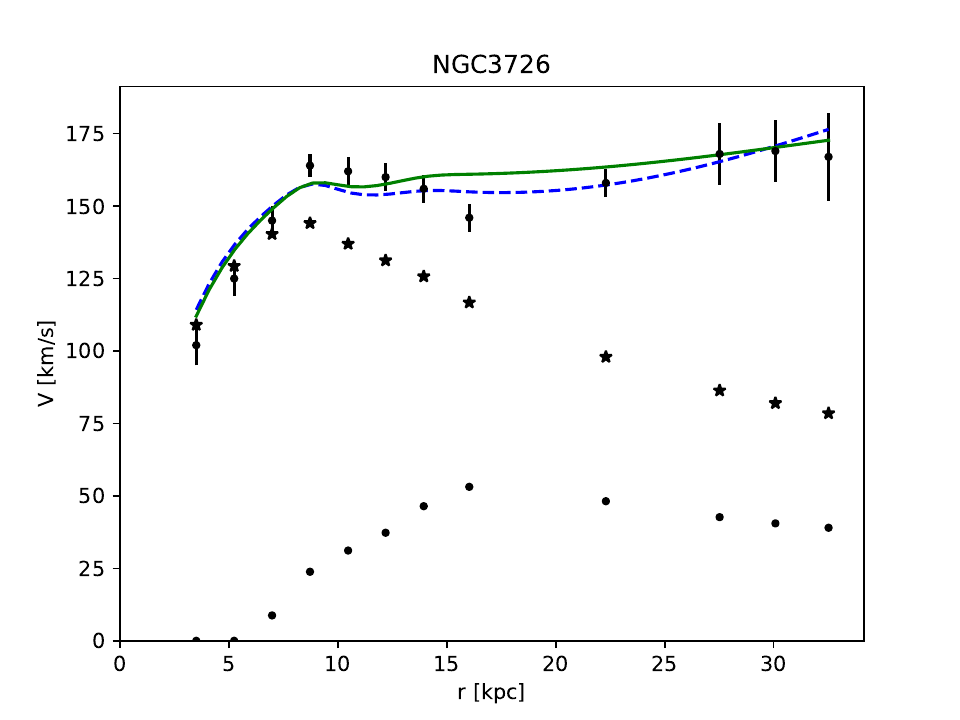} }  
  \subfloat[]{
        \label{fig:fit6}         
        \includegraphics[width=0.3\textwidth]{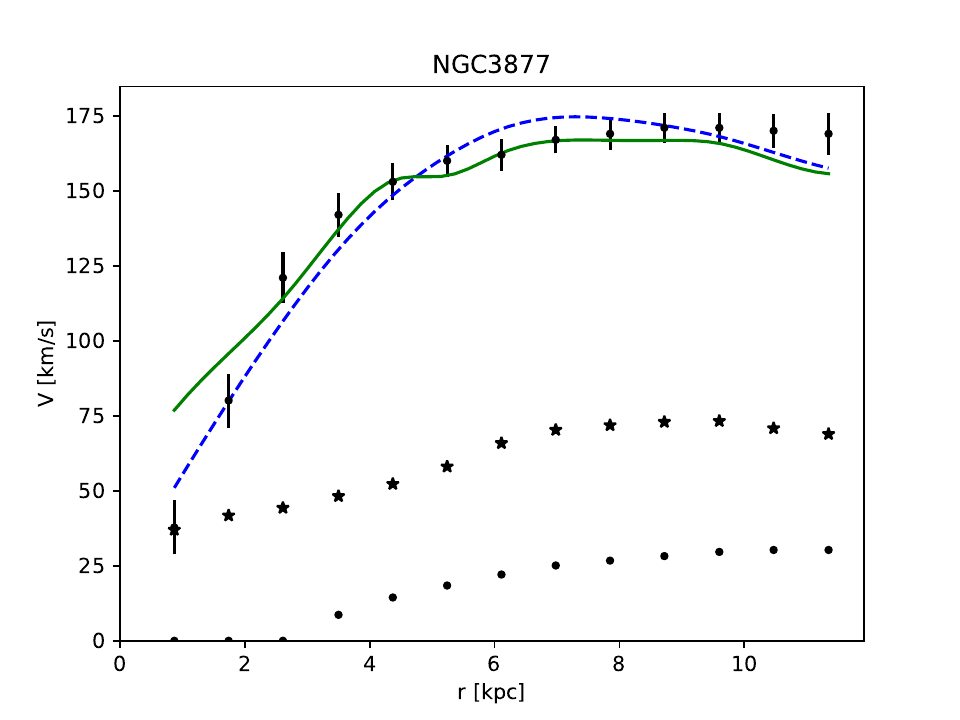} }
\\
   \subfloat[]{
        \label{fig:fit7}         
        \includegraphics[width=0.3\textwidth]{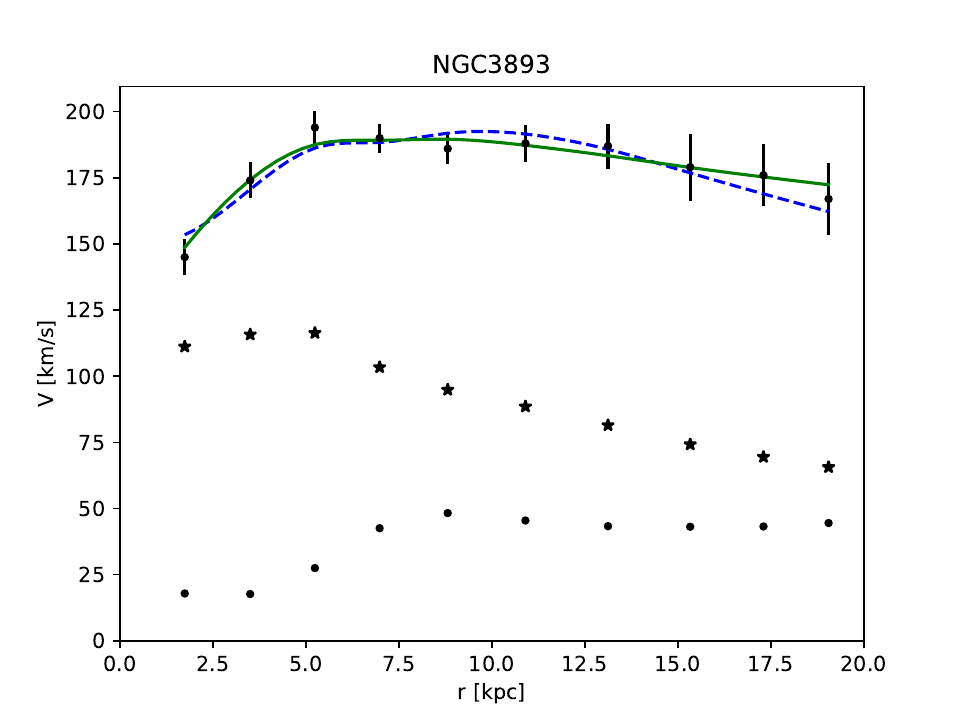} }   
   \subfloat[]{
        \label{fig:fit8}         
        \includegraphics[width=0.3\textwidth]{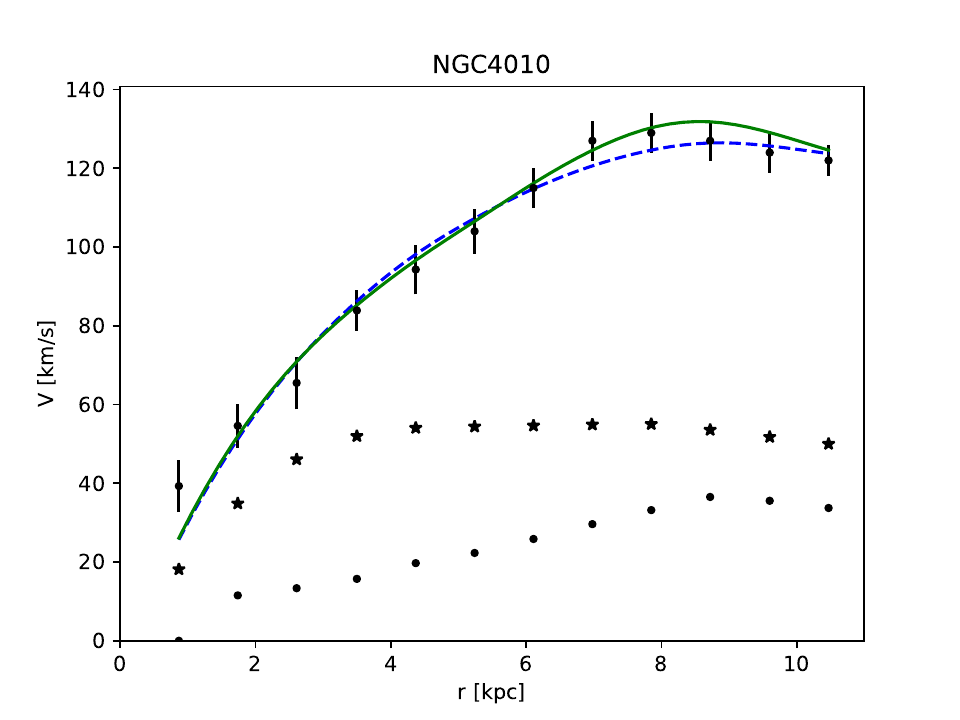} }  
  \subfloat[]{
        \label{fig:fit9}         
        \includegraphics[width=0.3\textwidth]{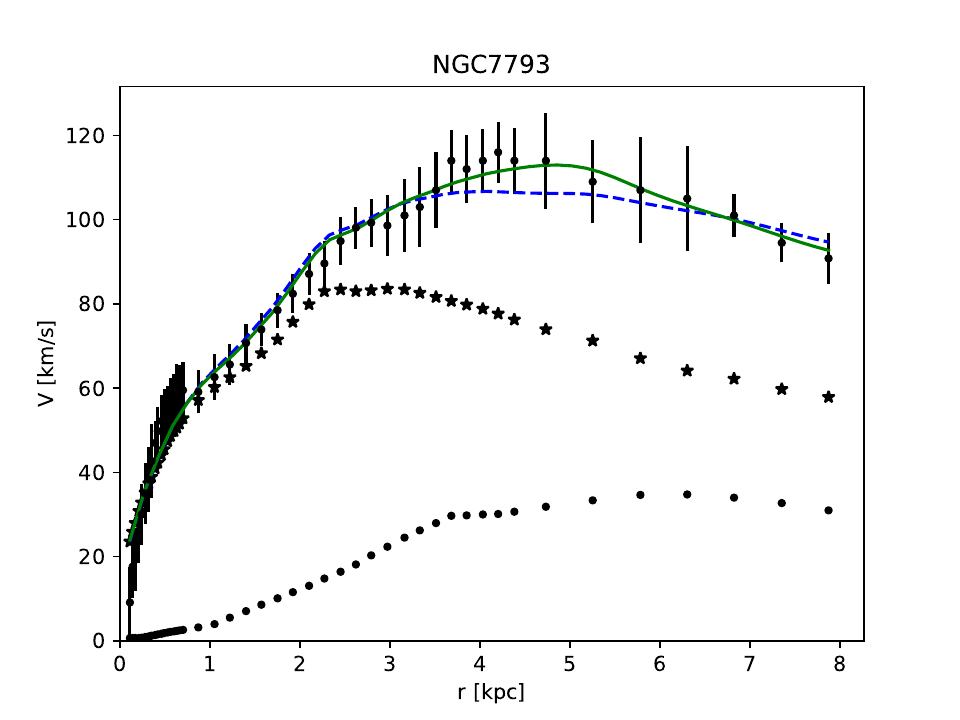} }

   \caption{Best fit of NC-like (dashed line) and Einasto (solid line) profiles over galaxy rotation velocity data (points with error bars), disk velocity (star markers), and gas velocity (circle markers). The total rotation velocity is obtained from Eq. \eqref{eq:Vtotal}.}
   \label{fig:bestfits}                
\end{figure*}

The Tables \ref{tab:ResultsNC} and \ref{tab:ResultsEIN} present the best fit values with their uncertainties at $68\%$ ($1\sigma$) confidence level (CL) for 
NC-like and Einasto respectively. Additionally, it is listed the reduced $\chi^2$ defined as
$\chi^2_{\mathrm{red}}=\chi^2/(N-p)$, where $N$ is the total number of data points and $p$ is the number 
of free model parameters. To compare the results of these models, we present for the Einasto case, the density evaluated at $r=0$ and $r_{-2}/2$ as an equivalent to parameter $\sqrt{\theta}$. Figure \ref{fig:bestfits} presents the best fits for NC-like (dashed line) and Einasto (solid line) profile over total velocity data, the disk (star markers) and gas (circle markers) components. Furthermore, we estimate the halo mass within a radius $r_{200}$ for each model (shown in the last column of the Tables \ref{tab:ResultsNC} and \ref{tab:ResultsEIN}), e.i., the mass contained in a radius which the density is 200 times the critical density of the universe. We find consistent values using Einasto model to those obtained by \cite{SPARC:2016,Li_2020} within $1\sigma$, except for NGC3726, NGC3867 which are deviated at $4.1\sigma$, $3.2\sigma$ respectively. When we compare our best fit values, we find consistent values within $1\sigma$,  for $r_s$ and $\log_{10}(\rho_{-2})$ respectively, and deviations up to $4.1\sigma$ for the Einasto index except for NGC2366, NGC2403, NGC3198 which obtain deviations larger than $5\sigma$.

To improve a statistical comparison of both models, we use the corrected Akaike information criterion (AICc) \cite{AIC:1974, Sugiura:1978, AICc:1989} defined as ${\rm AICc}= \chi^2_{min}+2k +(2k^2+2k)/(N-k-1)$ and the Bayesian information criterion (BIC) \cite{schwarz1978} defined as ${\rm BIC}=\chi^2_{min}+k\log(N)$. In the previous expressions, $k$ is the size of the parameter space and $N$ is the number of data points. The model with lower values of AICc and BIC is the one preferred by the data. In this context, the difference between the AICc value of a given model and the best one is denoted as $\Delta\rm{AICc}$. If $\Delta \rm{AICc}<4$, both models are supported by the data equally, if $4<\Delta\rm{AICc}<10$ the data still support the given model but less than the preferred one.
A value of $\Delta \rm{AICc}>10$ indicates that the data does not support the given model.
In contrast, $\Delta$BIC gives the evidence against a candidate model over the best model, which is the one with lower BIC value. Then,
a yield of $\Delta \rm{BIC}<2$ suggests that there is no evidence against the candidate model.
A value within $2<\Delta\rm{BIC}<6$ indicates that there is modest evidence against the candidate model. 
A strong evidence against the candidate model happens when $6<\Delta \rm{BIC}<10$, and 
a stronger evidence against is whether $\Delta \rm{BIC}>10$.
According to our AIC results shown in Table \ref{tab:ResultsBIC}, the galaxies NGC2366, NGC3893, and NGC4010 prefer both models equally, the galaxies NGC3521, NGC3726, and NGC7793 prefer NC-like over Einasto, and the galaxy NGC3877 does not support Einasto. Only the galaxies NGC2403 and NGC3198 do not support NC-like. 
Based on BIC, our results indicate that the galaxies NGC3893 and NGC4010 do not suggest a evidence against any model. The galaxies NGC2366 and NGC3726 suggest a modest evidence against Einasto model but a strong evidence using NGC3521 and NGC3877. In contrast, the sample NGC7793 gives a modest evidence against NC but a stronger evidence is observed using NGC2403 and NGC3198.

\begin{figure*}[h]
   \centering
   \subfloat[]{
        \label{fig:contour1}         
        \includegraphics[width=0.3\textwidth]{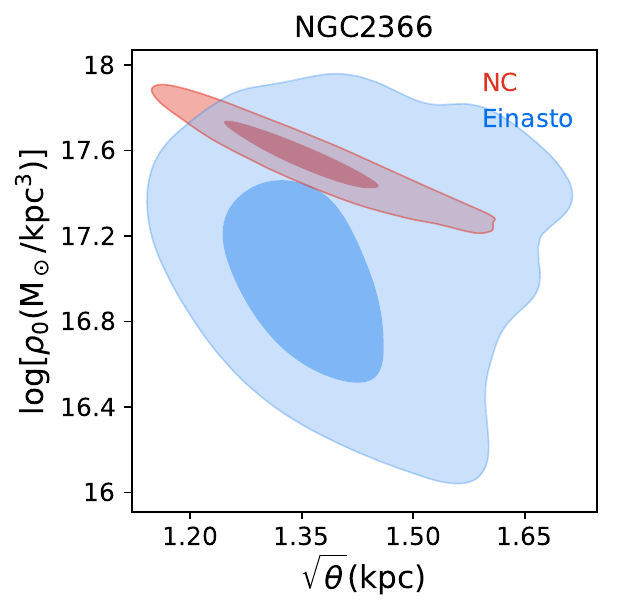} }           
   \subfloat[]{
       \label{fig:contour2}         
        \includegraphics[width=0.3\textwidth]{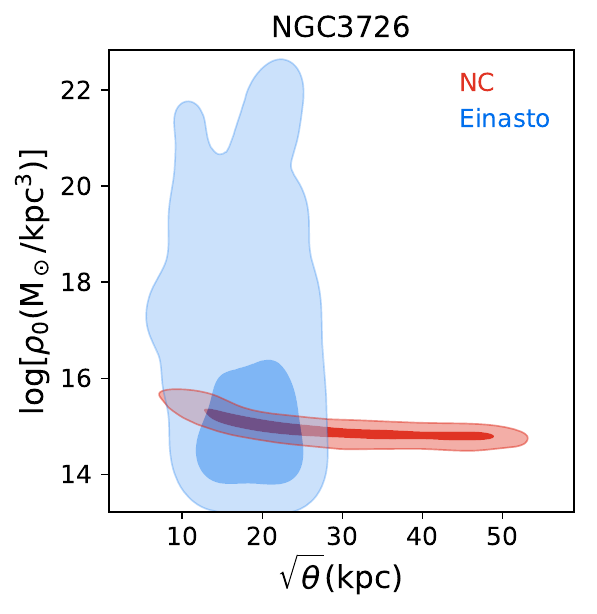} }
   \subfloat[]{
        \label{fig:contour3}         
        \includegraphics[width=0.3\textwidth]{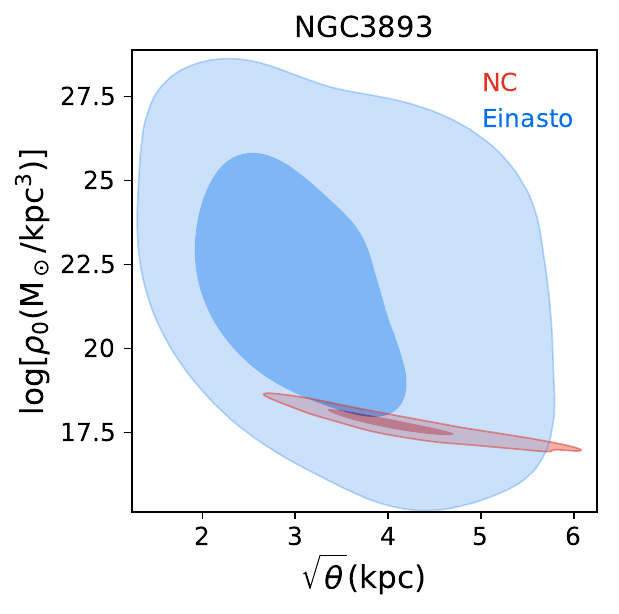} } 

   \caption{Contours at 1 and 2 $\sigma$ CL of NGC2366, NGC3521 and NGC7793 for NC-like and Einasto profiles in the parameters: central density $\rho_0$ and $\sqrt{\theta}$, ($r_{-2}/2$). The corresponding mean value of the spectral index in the Einasto model is $n\leq 1.3$. Darker region represent $1\sigma$ and lighter region is $3\sigma$}
   \label{fig:Contours2}                
\end{figure*}

\begin{figure*}[h]
   \centering
   \subfloat[]{
        \label{fig:contourEIN}         
        \includegraphics[width=0.4\textwidth]{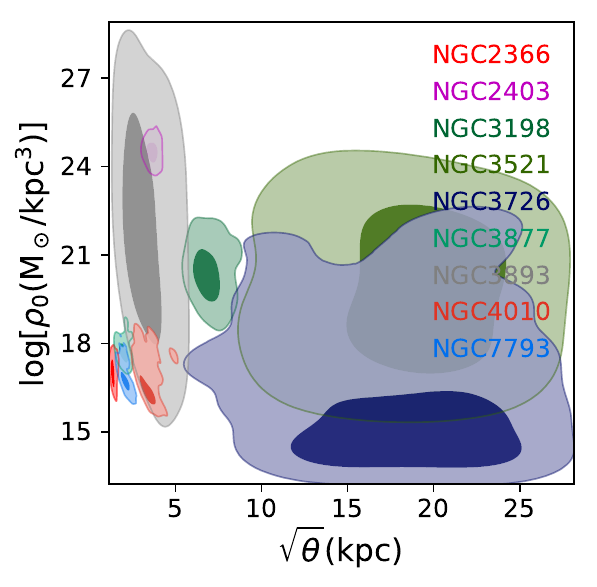} }           
   \subfloat[]{
       \label{fig:contourNC}         
        \includegraphics[width=0.4\textwidth]{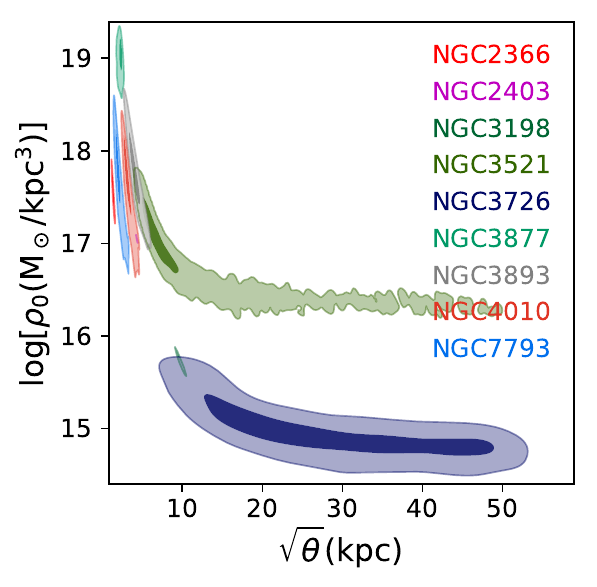} }

   \caption{Correlation of the parameter model ($\rho_0$ and $\sqrt{\theta}$ or $r_{-2}/2$) for: Einasto (left)  and NC-like (right). Darker region represent $1\sigma$ and lighter region is $3\sigma$}
   \label{fig:Contours3}                
\end{figure*}

Furthermore, Fig. \ref{fig:Contours2} displays the comparison of the space $(\rho_0$,$\sqrt{\theta})$ obtained by fitting the NGC2366, NGC3726 and NGC3893. These contours show consistent values for $n=1/2$ within up to $95\%\, (2\sigma)$ confidence level (CL) value which the NC-like model is recovered from Einasto. The contours represent $68\%\, (1\sigma)$ and 
$99.7\%\, (3\sigma)$ CL respectively.
On the other hand, we also present in Fig. \ref{fig:Contours3}, the correlations between $\log_{10}(\rho_0)$ and $\sqrt{\theta}$ for both models which is observed an uncorrelated between them, when Einasto is considered and also the uncertainties are considerable larger than these obtained using NC-like because the Einasto model has one more free parameter than NC-like. In this panel, contours at $1\sigma$ and $3\sigma$ are included.

\begin{table*}[h]
\begin{tabular}{llcccc}
\hline \hline
Galaxy  & $\chi^2_{\mathrm{red}}$ & $\rho_{0}$ $(10^{7}\mathrm{M}_{\odot}/\mathrm{kpc}^3)$ & $\sqrt{\theta}\, (\mathrm{kpc})$ &
$\Upsilon_d$ & $\log_{10}(M_{200}\,[\mathrm{M}_{\odot}])$\\
\hline \hline
NGC2366 & 0.17 & $4.27^{+0.48}_{-0.43}$ & $1.35^{+0.07}_{-0.06}$ & $0.29^{+0.13}_{-0.12}$ & $9.37^{+0.03}_{-0.03}$ \\ 
NGC2403 & 24.32 & $2.53^{+0.04}_{-0.03}$ & $4.33^{+0.03}_{-0.03}$ & $0.92^{+0.01}_{-0.01}$ & $10.66^{+0.01}_{-0.01}$ \\ 
NGC3198 & 4.42 & $0.67^{+0.03}_{-0.03}$ & $9.75^{+0.21}_{-0.20}$ & $0.92^{+0.01}_{-0.01}$ & $11.14^{+0.01}_{-0.01}$ \\ 
NGC3521 & 0.16 & $2.43^{+0.76}_{-0.67}$ & $6.95^{+2.79}_{-1.23}$ & $0.59^{+0.01}_{-0.01}$ & $11.26^{+0.30}_{-0.14}$ \\ 
NGC3726 & 2.01 & $0.30^{+0.08}_{-0.04}$ & $26.65^{+15.76}_{-10.39}$ & $0.67^{+0.03}_{-0.03}$ & $12.09^{+0.56}_{-0.54}$ \\ 
NGC3877 & 1.91 & $18.45^{+1.97}_{-2.03}$ & $2.31^{+0.10}_{-0.14}$ & $0.13^{+0.10}_{-0.07}$ & $10.71^{+0.06}_{-0.09}$ \\ 
NGC3893 & 0.74 & $5.20^{+1.49}_{-1.17}$ & $4.01^{+0.49}_{-0.42}$ & $0.53^{+0.05}_{-0.05}$ & $10.88^{+0.06}_{-0.06}$ \\ 
NGC4010 & 0.97 & $4.91^{+1.53}_{-1.34}$ & $3.10^{+0.40}_{-0.26}$ & $0.31^{+0.16}_{-0.15}$ & $10.52^{+0.04}_{-0.05}$ \\ 
NGC7793 & 0.61 & $5.28^{+1.56}_{-1.26}$ & $1.96^{+0.24}_{-0.18}$ & $0.75^{+0.05}_{-0.06}$ & $9.96^{+0.06}_{-0.06}$ \\
\hline \hline
\end{tabular}
\caption{For NC-like model, from left to right: Name of LSB galaxies of the SPARC, the reduced $\chi^2$, energy density, $\sqrt{\theta}$, $\Upsilon_d$ parameter, and $M_{200}$ is the DM mass within a radius $r_{200}$ which the density is $\rho_{200}(r_{200})$ (200 times the critical density of the universe).}
\label{tab:ResultsNC}
\end{table*}


\begin{table*}[h]
\begin{tabular}{llccccc}
\hline \hline
Galaxy  & $\chi^2_{\mathrm{red}}$ & $\rho_{-2}\,e^{2n}$ $(10^{7} \mathrm{M}_{\odot}/\mathrm{kpc}^3)$ & $r_{-2}/2\, (\mathrm{kpc})$ & $n$ & $\Upsilon_d$ & $\log_{10}(M_{200}\,[\mathrm{M}_{\odot}])$ \\
\hline \hline
NGC2366 & 0.19 & $2.46^{+0.91}_{-0.67}$ & $1.36^{+0.08}_{-0.07}$ & $0.58^{+0.17}_{-0.16}$ & $0.24^{+0.19}_{-0.17}$ & $9.35^{+0.03}_{-0.04}$  ($9.38\pm 0.07$) \\ [.8ex]
NGC2403 & 9.32 & $4352.92^{+1072.95}_{-862.10}$ & $3.66^{+0.19}_{-0.22}$ & $4.47^{+0.12}_{-0.11}$ & $0.25^{+0.03}_{-0.03}$ & $11.35^{+0.03}_{-0.03}$ ($11.38\pm 0.06$) \\ [.8ex]
NGC3198 & 1.03 & $67.10^{+53.90}_{-29.86}$ & $6.84^{+0.54}_{-0.51}$ & $2.86^{+0.30}_{-0.26}$ & $0.49^{+0.04}_{-0.05}$ & $11.49^{+0.05}_{-0.05}$ ($11.44\pm 0.05$) \\ [.8ex]
NGC3521 & 0.33 & $45.19^{+260.89}_{-40.02}$ & $19.44^{+3.29}_{-3.76}$ & $3.28^{+1.08}_{-1.26}$ & $0.55^{+0.03}_{-0.02}$ & $12.39^{+0.16}_{-0.20}$ ($12.26\pm 0.85$) \\ [.8ex]
NGC3726 & 2.62 & $0.30^{+1.38}_{-0.15}$ & $18.59^{+4.32}_{-4.75}$ & $0.88^{+0.93}_{-0.57}$ & $0.62^{+0.05}_{-0.08}$ & $11.69^{+0.29}_{-0.42}$ ($13.48\pm 0.32$) \\ [.8ex]
NGC3877 & 3.6 & $4.82^{+1.02}_{-0.92}$ & $1.98^{+0.18}_{-0.16}$ & $0.01^{+0.11}_{-0.01}$ & $0.51^{+0.04}_{-0.05}$ & $10.17^{+0.08}_{-0.08}$ ($10.55\pm 0.09$) \\ [.8ex]
NGC3893 & 0.34 & $291.33^{+3496.21}_{-264.45}$ & $2.97^{+1.05}_{-0.61}$ & $2.55^{+1.33}_{-1.04}$ & $0.30^{+0.12}_{-0.12}$ & $11.24^{+0.23}_{-0.20}$ ($11.02\pm 0.39$) \\ [.8ex]
NGC4010 & 0.99 & $1.32^{+0.81}_{-0.29}$ & $3.36^{+0.29}_{-0.30}$ & $0.18^{+0.24}_{-0.13}$ & $0.47^{+0.11}_{-0.16}$ & $10.38^{+0.09}_{-0.06}$ ($10.47\pm 0.68$) \\ [.8ex]
NGC7793 & 0.46 & $1.82^{+0.41}_{-0.32}$ & $2.08^{+0.17}_{-0.16}$ & $0.10^{+0.13}_{-0.07}$ & $0.76^{+0.04}_{-0.04}$ & $9.88^{+0.05}_{-0.06}$  ($9.87\pm 0.56$) \\ [.8ex]
\hline \hline
\end{tabular}
\caption{For Einasto model, from left to right: Name of LSB galaxies of the SPARC catalogue under study, the reduced $\chi^2$, the density at $r=0$, $r_0=r_{-2}/2$, the spectral index $n$, $\Upsilon_d$ parameter and $M_{200}$ is the DM mass within a radius $r_{200}$ which the density is $\rho_{200}(r_{200})$ (200 times the critical density of the universe). The values in parentheses in the last column are those halo mass reported in \cite{SPARC:2016}. }
\label{tab:ResultsEIN}
\end{table*}


\begin{table*}[h]
\begin{tabular}{ccccccc}
\hline \hline
Galaxy  & AICc(NC-like) & AICc(Einasto) & $|\Delta$AIC$|$& BIC(NC-like) & BIC(Einasto) & $|\Delta$BIC$|$\\
\hline \hline
NGC2366 & $10.9$ 	 & $14.0$ 	 & $3.1$ 	 & $13.6$ 	 & $17.1$ 	 & $3.5$    \\ [.8ex]
NGC2403 & $1708.5$ 	 & $651.5$ 	 & $1057.0$  & $1715.0$  & $660.1$ 	 & $1054.9$ \\ [.8ex]
NGC3198 & $183.3$ 	 & $49.1$ 	 & $134.2$ 	 & $187.9$ 	 & $55.0$ 	 & $132.9$  \\ [.8ex]
NGC3521 & $12.7$ 	 & $21.2$ 	 & $8.5$ 	 & $17.2$ 	 & $26.9$ 	 & $9.8$    \\ [.8ex]
NGC3726 & $27.1$ 	 & $34.7$ 	 & $7.6$ 	 & $25.5$ 	 & $30.9$ 	 & $5.3$    \\ [.8ex]
NGC3877 & $27.8$ 	 & $45.4$ 	 & $17.6$ 	 & $26.8$ 	 & $42.7$ 	 & $15.9$   \\ [.8ex]
NGC3893 & $15.2$ 	 & $18.1$ 	 & $2.9$ 	 & $12.1$ 	 & $11.3$ 	 & $0.8$    \\ [.8ex]
NGC4010 & $17.8$ 	 & $21.7$ 	 & $3.9$ 	 & $16.2$ 	 & $17.9$ 	 & $1.7$    \\ [.8ex]
NGC7793 & $32.9$ 	 & $28.4$ 	 & $4.5$ 	 & $37.9$ 	 & $34.8$ 	 & $3.1$    \\ [.8ex]
\hline \hline
\end{tabular}
\caption{AIC and BIC for NC-like and Einasto. $\Delta$AIC ($\Delta$BIC) yield value is the difference between AIC (BIC) model values of AIC for each galaxy.}
\label{tab:ResultsBIC}
\end{table*}

\section{Discussion} \label{Con}

This manuscript was devoted to the comparison of the well-known Einasto density profile to describe DM halos at galactic scale with a particular profile obtained when the Einasto index is $n=1/2$ and named as NC-like.  The NC-like model, is motivated by the idea of a noncommutative space-time coming from string theory. Then, the strategy followed was to confront the total rotation curve using both models with a sample of galaxies provided by SPARC catalogue \cite{SPARC:2016} by performing a Bayesian MCMC procedure. To model the rotation curves, we considered the gas, disk and dark components, which the latter corresponds to NC-like or Einasto model and the first two are provided by data. 
According to our results, for some galaxies (NGC3521, NGC3726, and NGC7793) we found a better preference of the NC-like over Einasto model but other galaxies (NGC2403 and NGC3198) do not support NC. Although the galaxies NGC2366, NGC3893, and NGC4010 prefer both models equally, NC-like has the advantage that allows to obtain best fitting values with an uncertainty lower than those obtained by Einasto because the parameter $n=1/2$ is fixed. We compare the contours obtained for NGC3521, NGC3726, and NGC7793 at $1\sigma$ and $3\sigma$ showing consistent regions because the Einasto indexes obtained are in agreement with the value for NC-like. Additionally, from Tables \ref{tab:ResultsNC} and \ref{tab:ResultsEIN}, we conclude that masses in NC and Einasto are at the same order of magnitude, even, some Einasto masses are larger that the NC case. At least for the case $n=1/2$ (Einasto case), both masses are in good agreement. Indeed, this result is expected because NC could be consider as a particular case of Einasto density profile.

From the results presented in Table \ref{tab:ResultsNC} we can observe that NC length is not an invariant quantity and therefore is not a fundamental structure. However, it is expected that the $\theta$ factor is constructed by quantum cells that emerges from NC theory. The presence of the density profile (Eq. \eqref{Gauss}) in galaxies is the emergence of non-commutative quantum properties of space-time and may be an indirect evidence of the granular structure. It is important to remark that we are hypothesizing about the quantum structure and the mathematical support is not the aim of the present paper.

This work motivates further studies with more statistics and include new observables to allow discriminate between the NC-like model and others (Einasto, piso, etc) and also find correlations between NC properties and the characteristic of the galaxies, see for instance \cite{Hernandez-Almada:2017mtm,Li_2020}. Note that a joint analysis does not strengthen our results, since the parameters are not invariant and they depend on characteristics such as the size of the galaxy. Additionally, it could be interesting to perform a statistical test to compare the NC-like model with other densities using alternative criteria such as Bridge criterion presented in \cite{Ding:2018}. Finally, we expect that accumulative terms of NC-like can have macroscopic repercussions that might be explored as a possible solution to fundamental problems like dark matter or dark energy, having always in mind that the NC-like profile has theoretical foundations of great importance and repercussion.

\begin{acknowledgments}
We thank the anonymous referees for thoughtful remarks and suggestions. A.H.A. thanks to SNI-M\'exico for partial financial support. M.A.G.-A. acknowledges support from SNI-M\'exico, CONACyT research fellow, CONICYT REDES (190147), COZCyT and Instituto Avanzado de Cosmolog\'ia (IAC).
\end{acknowledgments}

\bibliography{librero0}

\end{document}